# Birefringence of Muscovite Mica Plate: Temperature Effect in the Ultraviolet and Visible Spectrum

Xu Zhang [1,2*], Fuquan Wu [2], Limei Qi [2], Xia Zhang [2], Dianzhong Hao [2]

*1. School of Physics Science and Engineering, Tongji University, Shanghai, 200092, China*
*2. Shandong Provincial Key Laboratory of Laser Polarization and Information Technology, College of Physics and Engineering, Qufu Normal University, Qufu, 273165, China*
*[*]Corresponding author: zhangxuyq@163.com*

**Abstract** We developed a method to measure the phase retardation and birefringence of muscovite mica plate in the temperature range of 223K to 358K within the spectrum of 300 to 700 nm. The phase retardation data is gained through the standard transmission ellipsometry using spectroscopic ellipsometer. With the phase retardation and thickness of the mica plate we can calculate its birefringence dispersion. Our results give abundant phase retardation and birefringence data of muscovite mica in the ultraviolet and visible spectrum from 223K to 358K. From the experimental data, the phase retardation and birefringence will drop down at the fixed wavelength when the temperature rises. The accuracy of the birefringence of mica plate is better than $3.5 \times 10^{-5}$.

*Keywords :* Phase retardation, Birefringence, Muscovite mica wave plate, Temperature effect
*OCIS codes:* (120.2130) Ellipsometry and polarimetry; (120.5050) Phase measurement, (120.0120) Instrumentation, measurement, and metrology.

## 1 Introduction

Muscovite mica is widely used in polarization technology [1,2] and as substrate in technological and biological applications [3-5]. Because of its low thickness mica is the ideal crystal to manufacture the zero order wave plate which is not sensitive to the wavelength [6]. On the contrary the designed multiple-order plate can only work at a single wavelength because a slight shift from the designed wavelength will cause a large variation to the phase retardation of the plate [7,8]. But the birefringence dispersion of mica plate is not always the same as to different samples [1,9-12] which makes the determination of different mica sample's birefringence dispersion essential.

In this work, we develop a method to determine the phase retardation and birefringence dispersion of the mica plate between 300 and 700 nm. Before the measurement we coated the mica plate to increase the transmittance because there are multiple coherent reflections inner the mica plate which have great influence on the phase retardation (shown in part 2.2). Using the spectroscopic ellipsometer we gain the phase retardation of the mica plate. With the thickness and phase retardation of the mica plate we can calculate its birefringence dispersion. Since the birefringence dispersion of mica plate under different temperatures may be used in its applications, we measured the birefringence dispersion in the temperature range of 223K to 358K.

## 2 Measuring principle

### 2.1 Ellipsometry

The standard ellipsometry includes the reflection ellipsometry and the transmission ellipsometry. Ellipsometric angles $\psi$ and $\Delta$ are two important parameters in the ellipsometry method. As to the isotropic medium, the relationship between the ellipsometry parameters and the reflection (or transmission) matrix $\begin{bmatrix} j_p & 0 \\ 0 & j_s \end{bmatrix}$ is [13,14]:

$$\rho = \frac{j_p}{j_s} = \tan\psi \exp(i\Delta) . \qquad (1)$$

As to the reflection ellipsometry, $j_p$ and $j_s$ denote the p- and s-polarized complex reflection coefficients ($j_{p,s} = r_{p,s}$). To transmission ellipsometry, they denote the p- and s-polarized complex transmission coefficients ($j_{p,s} = t_{p,s}$).

But to the anisotropic medium, the generalized ellipsometry should be used. The reflection (or transmission) matrix is not diagonalized, it is $\begin{bmatrix} j_{pp} & j_{sp} \\ j_{ps} & j_{ss} \end{bmatrix}$. The relationship between the ellipsometry parameters and this matrix is [13,14]:

$$\begin{cases} \rho_{ps} = \dfrac{j_{ps}}{j_{pp}} = \tan\psi_{ps} \exp(i\Delta_{ps}) \\ \rho_{sp} = \dfrac{j_{sp}}{j_{ss}} = \tan\psi_{sp} \exp(i\Delta_{sp}) \\ \rho_{pp} = \dfrac{j_{pp}}{j_{ss}} = \tan\psi_{pp} \exp(i\Delta_{pp}) \end{cases} \qquad (2)$$

The reflection (or transmission) matrix can be diagonalized as $\begin{bmatrix} j_p & 0 \\ 0 & j_s \end{bmatrix}$ when the optic axis is parallel or vertical to the incident plane [14,15]. Under this condition, the standard ellipsometry can still be used to the anisotropic medium.

As to the standard transmission ellipsometry,

$$\rho_t = \frac{t_p}{t_s} = \tan\psi_t \exp(i\Delta_t) , \qquad (3)$$

where $t_p = |t_p|\exp(i\Delta_{tp})$ ($\Delta_{tp}$ is the phase shift of the p vibration), $t_s = |t_s|\exp(i\Delta_{ts})$ ($\Delta_{ts}$ is the phase shift of the s vibration), $\psi_t = \tan^{-1}|\rho_t|$, $\Delta_t = \Delta_{tp} - \Delta_{ts}$ is the phase retardation between the p vibration and s vibration.

When a linearly polarized light is incident normally upon a mica wave plate, it will be decomposed into o light (s vibration) and e light (p vibration). The refractive index is $n_o$ and $n_e$ respectively. Because the two lights have different speeds ($v_e > v_o$) in the mica plate, after they passed through the mica plate with the thickness of d the phase retardation $\Delta$ between them is [1]:

$$\Delta = -2\pi(n_e - n_0)d/\lambda , \qquad (4)$$

where $(n_e - n_0)$ is the birefringence at the wavelength $\lambda$. Muscovite mica is biaxial crystal. In the manufacture of mica plate, from the property of the cleavage plane we know $n_e = n_\beta$, $n_o = n_\gamma$ [1,16,17]. Then the birefringence of the mica wave plate is $n_e - n_0 = n_\beta - n_\gamma$. From the above discussion, the standard transmission ellipsometry can be used to measure the phase retardation of the mica wave plate.

## 2.2 Multiple coherent reflections in mica plate [18]

There are multiple coherent reflections within the mica plate because of its little thickness and excellent flatness which is shown in Fig. 1. The fast and slow axes of the mica plate are set as the $x$ and $y$ axis separately and the propagation direction is set as $z$ axis. I is the incident light intensity, $R_1, R_2, R_3$ ……. are the reflected light and $T_1, T_2, T_3$ ……. are the transmitted light.

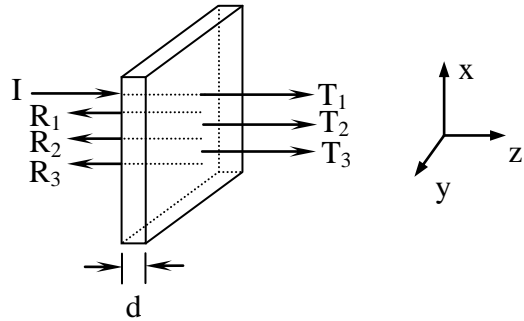

Fig.1 Multiple coherent reflections in mica wave plate, $d$ is the width of the mica plate.

As we all know the Jones matrix of the mica wave plate is $\begin{bmatrix} \exp(-i\delta_x) & 0 \\ 0 & \exp(-i\delta_y) \end{bmatrix}$ without the consideration of the multiple reflections within it. Considering the multiple reflections, the Jones matrix should be (shown in Appendix A):

$$J_m = \begin{bmatrix} \dfrac{t_x^2 \exp(-i\delta_x)}{1+r_x^2 \exp(-i2\delta_x)} & 0 \\ 0 & \dfrac{t_y^2 \exp(-i\delta_y)}{1+r_y^2 \exp(-i2\delta_y)} \end{bmatrix}, \qquad (5)$$

where $r_x = \dfrac{n - n_x}{n + n_x}$, $t_x = \dfrac{2n}{n + n_x}$, $\delta_x = \dfrac{2\pi n_x d}{\lambda}$, $r_y = \dfrac{n - n_y}{n + n_y}$, $t_y = \dfrac{2n}{n + n_y}$, $\delta_y = \dfrac{2\pi n_y d}{\lambda}$. Then $J_m$ can be set as

$$J_m = \psi_x \exp(i\Delta_x) \begin{bmatrix} 1 & 0 \\ 0 & \psi \exp(i\Delta) \end{bmatrix}. \qquad (6)$$

The final form of equation (6) is the Jones matrix of the mica plate considering the multiple reflections within it. From this Jones matrix we can see $\psi$ is the complex amplitude ratio of the transmitted light at the $x$ and $y$ direction and $\Delta$ is the phase retardation between them ($\Delta = \Delta_y - \Delta_x$). From calculation we can gain (shown in Appendix A):

$$\Delta = \arctan\left(\frac{(1-r_{1y}^2)(1+r_{1x}^2)\tan\delta_y - (1-r_{1x}^2)(1+r_{1y}^2)\tan\delta_x}{(1+r_{1y}^2)(1+r_{1x}^2) + (1-r_{1y}^2)(1-r_{1x}^2)\tan\delta_x\tan\delta_y}\right). \quad (7)$$

In order to display the effect of the multiple reflections within mica plate, we calculated the phase retardation curve of a mica plate (a quarter wave plate at 633nm) in the spectrum of 550-700nm. The reflection index of mica wave plate before coated is about 5.18%. Then from equation (7) we can gain Fig. 2(a). The red line is the result without the consideration of multiple reflections. If we set the reflection index as 1.1%, we can gain Fig. 2(b). It is quite evident that the oscillations will be reduced a lot with the decrease of the reflection index which enables us to gain more accurate phase retardation data. So we should coat the mica wave plate before we use the standard transmission ellipsometry to measure its phase retardation and birefringence.

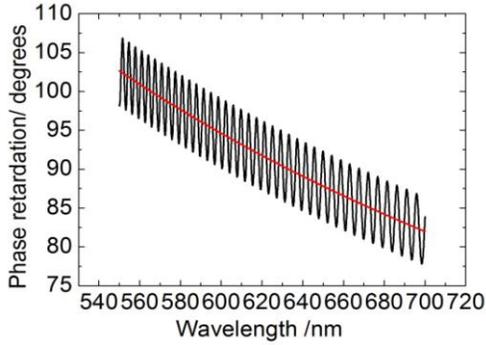 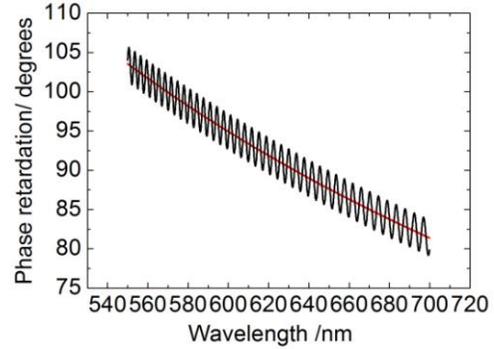

Fig.2 (a) The phase retardation of the mica quarter wave plate (reflection index =5.18%)

Fig.2 (b) The phase retardation of the mica quarter wave plate (reflection index =1.1%)

## 3 Experiment

We evaporate broadband anti-reflective film on the mica wave plate. The film coefficient is Sub/Al$_2$O$_3$^1/WD10^2/MgF$_2$^1/Air. The anti-reflective film can increase the transmission in the ultraviolet and visible spectrum and suppress the multiple coherent reflections within the plate (shown in part 2.2).

### 3.1 Phase retardation measurement

The UVISEL spectroscopic phase modulated ellipsometer (USPME, shown in Fig. 3(a)) made by the French Jobin Yvon corporation is used in the experiment to measure the phase retardation of the mica wave plate. $L$ is laser, $P$ the polarizer, $S$ the measured mica wave plate, $M$ the modulator, $A$ the analyzer and $D$ the detector.

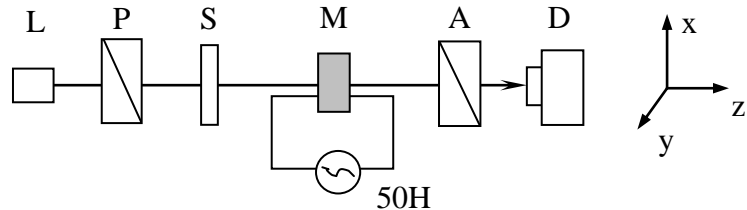

Fig. 3(a) Optical system for measuring the phase retardation and birefringence of mica plate.

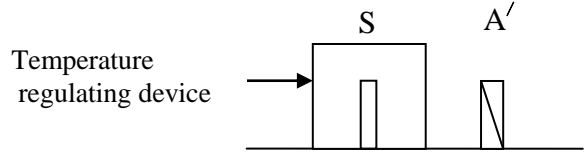

Fig. 3(b) Temperature regulating device on the sample table

As to the USPME, the polarizer $P$, modulator $M$ and analyzer $A$ have two configurations [19]. Configuration I is $P-M=45°$, $M=0°$, $A=45°$ and configuration II is $P-M=45°$, $M=45°$, $A=45°$. In configuration I, the final detected intensity collected by the detector is

$$I(t) = I[I_0 + I_s \sin\delta(t) + I_c \cos\delta(t)], \quad (8)$$

with

$$I_0 = 1, \quad I_s = \sin 2\psi \sin\Delta, \quad I_c = \sin 2\psi \cos\Delta. \quad (9)$$

While in configuration II, the final detected intensity collected by the detector is

$$I(t) = I[I_0 + I_s \sin\delta(t) + I_c \cos\delta(t)], \quad (10)$$

with

$$I_0 = 1, \quad I_s = \sin 2\psi \sin\Delta, \quad I_c = \cos 2\psi. \quad (11)$$

In configuration I we can measure $\Delta$ accurately over the full $0°-360°$ range (theoretical equations (9) give $\tan\Delta$ and $\sin 2\psi$). But when $\psi$ is near $45°$ we will suffer from $\psi$ accuracy problems. As to the configuration II, we can measure $\psi$ accurately over the full $0°-90°$ range, but suffer from $\Delta$ indetermination in the range $90°-270°$. Since we want to gain the precise phase retardation $\Delta$ of the mica plate, we use configuration I in the experiment. We developed the following experiment procedures:

Step 1, Polarizer $P$, modulator $M$ and analyzer $A$ are aligned in orientations with $P=0°$, $M=0°$, $A=45°$. Then place another analyzer $A'$ on the sample table shown in Fig. 3(b). Adjust $A'$ to make $P$ and $A'$ crossed. (We add another analyzer $A'$ because the angles of $M$ and $A$ are fixed in configuration I.)

Step 2, Place the measured mica plate between $P$ and $A'$, adjust the plate until minimum intensity (within the measurement noise level) is detected which means the optic axis of the measured plate is parallel or vertical to the transmission axis of the polarizer $P$. In the mean time the first harmonic component of the modulated intensity $R\omega$ and the second harmonic component $R2\omega$ should be zero (within the measurement noise level) which enables us to use the ellipsometric angles $\psi$ and $\Delta$ in the standard ellipsometric configurations to measure a biaxially anisotropic slab (the mica wave plate) [15].

Step 3, Set $P=90°$, then adjust $A'$ to make $P$ and $A'$ crossed again. If the first and the second harmonic components of the modulated intensity are still zero (within the measurement noise level), then the measured plate is vertical to the incident light.

Step 4, Take $A'$ away and adjust $P=45°$ (configuration I). Install the temperature regulating device on the sample table shown in Fig. 3(b). We adjust the temperature from 223K to 293K as to mica plate sample 1 with a temperature interval of 10K and 298K to 358K to mica plate sample 2 also with a temperature interval of 10K. Set the spectral range ($300-700$ nm) and wavelength interval of the UVISEL ellipsometer, the retardation of the wave plate can be gained from the outputted data.

Step 5, Then adjust $P = -45°$ to repeat the measurement in step 4. This step is used to check the p and s transmissions are same or not which can be used to judge the standard transmission ellipsometry can be applied or not. If p and s transmissions are different, there must be miscuts in the preparation of the wave plate. Then analysis schemes must invoke the full anisotropic generalized ellipsometry algorithm [14]. From our experiment results the p and s transmissions agree well with each other.

We set the wavelength interval in this experiment as 10 nm. The interval can be set as 5 nm, 1 nm and so on. So we can gain more data of the phase retardation when the interval is smaller.

### 3.2 Thickness measurement and birefringence calculation

In order to gain the birefringence of the mica plate, we should know the thickness of the plate from equation (4). Our mica plate samples were measured at different points by a high precise digital micrometer with a resolution ratio of 0.1μm at the temperature of 293K. The mean value of the sample 1 is 33.8 μm and 33.1 μm as to sample 2. From the distribution of thickness values for the two mica plates, the uncertainty in thickness values is about 0.2 μm.

Since the thickness of the mica plate $d_T$ at temperature T is [20]:

$$d_T = d_0[1 + 7.37 \times 10^{-6} \times (T - T_0)], \qquad (12)$$

where $d_0$ is the thickness of the mica plate at the temperature of 293K, $T_0$ is 293K. From equation (4) and (12), the birefringence of the mica plate at temperature T is:

$$(n_e - n_0)_T = -\frac{\lambda \Delta}{2\pi d_T}. \qquad (13)$$

Based on this equation we can study the temperature effect of mica plate's birefringence.

## 4 Results and discussions

### 4.1 Results analysis

The phase retardation of muscovite mica plate sample 1 is shown in Fig. 4 (a). Compared with Fig. 2, the oscillations have been eliminated by the coating of the anti-reflective film which enhances the accuracy of our phase retardation data. From Fig. 4 (a) when the temperature rises, the phase retardation drops down at the fixed wavelength. Since the thickness of mica plate will increase due to the heat expansion so the birefringence should decrease with the rise of the temperature at the fixed wavelength which is verified by Fig. 4(b). So the effect of the birefringence is the main factor in the temperature effect of the phase retardation of muscovite mica plate.

With the phase retardation data, we can calculate the birefringence of mica from equation (13) in the temperature range of 223K to 293K as to sample 1. The birefringence data of sample 1 are listed in Table 1. From Fig. 4(b) (drawn from the data in Table 1) we can find the birefringence decreases with the rise of the temperature at the fixed wavelength as mentioned above. When the temperature is fixed, the birefringence will go up with the increase of the wavelength. Oscillations of birefringence exist at some wavelengths which is in accordance with the data in reference [1].

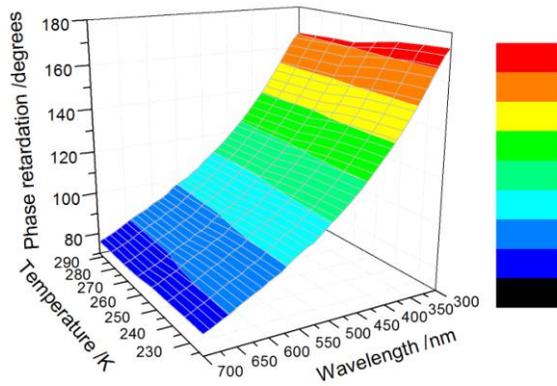
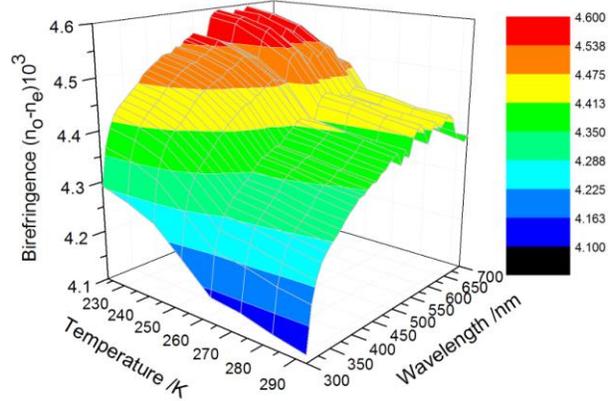

Fig.4 (a) Phase retardation of the mica plate of thickness 33.8μm

Fig.4(b) Birefringence of the mica plate of thickness 33.8μm

The phase retardation of muscovite mica plate sample 2 is shown in Fig. 5(a). Its birefringence data are listed in Table 2 and shown in Fig. 5(b). From the figures we can see the variety law of phase retardation and birefringence of sample 2 is same to sample 1.

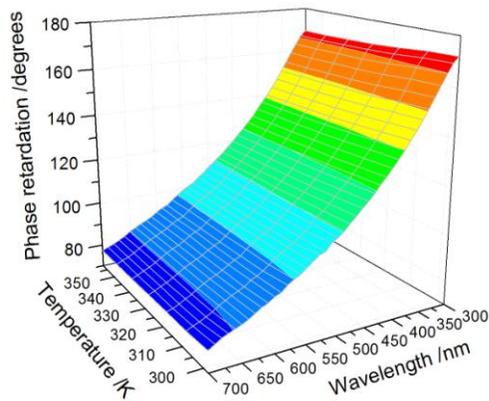
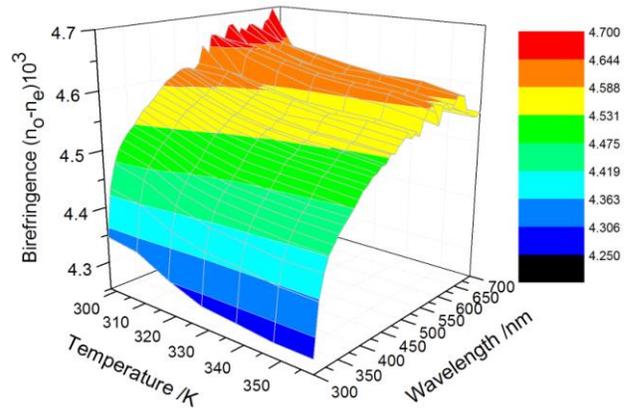

Fig.5 (a) Phase retardation of the mica plate of thickness 33.1μm

Fig.5 (b) Birefringence of the mica plate of thickness 33.1μm

Table 1 Birefringence $-(n_e - n_o) \times 10^3$ of muscovite mica plate sample 1 of thickness 33.8μm from 223K to 293K.

| $\lambda$ (nm) | 223K | 233K | 243K | 253K | 263K | 273K | 283K | 293K |
| --- | --- | --- | --- | --- | --- | --- | --- | --- |
| 300 | 4.291 | 4.274 | 4.250 | 4.201 | 4.148 | 4.141 | 4.126 | 4.115 |
| 310 | 4.366 | 4.353 | 4.330 | 4.281 | 4.231 | 4.222 | 4.207 | 4.198 |
| 320 | 4.401 | 4.389 | 4.366 | 4.320 | 4.267 | 4.259 | 4.245 | 4.235 |
| 330 | 4.425 | 4.412 | 4.390 | 4.343 | 4.291 | 4.283 | 4.271 | 4.261 |
| 340 | 4.439 | 4.425 | 4.404 | 4.357 | 4.304 | 4.297 | 4.284 | 4.274 |
| 350 | 4.447 | 4.436 | 4.414 | 4.369 | 4.316 | 4.307 | 4.295 | 4.285 |

| λ (nm) | | | | | | | | |
|---|---|---|---|---|---|---|---|---|
| 360 | 4.458 | 4.445 | 4.423 | 4.378 | 4.324 | 4.315 | 4.303 | 4.293 |
| 370 | 4.467 | 4.454 | 4.433 | 4.390 | 4.334 | 4.325 | 4.311 | 4.303 |
| 380 | 4.476 | 4.465 | 4.444 | 4.402 | 4.340 | 4.334 | 4.320 | 4.314 |
| 390 | 4.490 | 4.477 | 4.456 | 4.412 | 4.352 | 4.348 | 4.335 | 4.325 |
| 400 | 4.498 | 4.484 | 4.466 | 4.423 | 4.362 | 4.356 | 4.343 | 4.332 |
| 410 | 4.506 | 4.494 | 4.473 | 4.431 | 4.371 | 4.363 | 4.350 | 4.342 |
| 420 | 4.512 | 4.501 | 4.478 | 4.439 | 4.373 | 4.364 | 4.354 | 4.344 |
| 430 | 4.520 | 4.506 | 4.485 | 4.446 | 4.379 | 4.371 | 4.359 | 4.351 |
| 440 | 4.523 | 4.512 | 4.489 | 4.452 | 4.384 | 4.377 | 4.367 | 4.358 |
| 450 | 4.523 | 4.512 | 4.492 | 4.452 | 4.383 | 4.376 | 4.366 | 4.358 |
| 460 | 4.533 | 4.523 | 4.504 | 4.463 | 4.395 | 4.389 | 4.376 | 4.369 |
| 470 | 4.532 | 4.524 | 4.505 | 4.466 | 4.403 | 4.394 | 4.382 | 4.372 |
| 480 | 4.545 | 4.532 | 4.511 | 4.474 | 4.410 | 4.401 | 4.389 | 4.382 |
| 490 | 4.548 | 4.535 | 4.520 | 4.479 | 4.418 | 4.408 | 4.399 | 4.389 |
| 500 | 4.556 | 4.548 | 4.524 | 4.488 | 4.422 | 4.418 | 4.406 | 4.393 |
| 510 | 4.548 | 4.540 | 4.520 | 4.483 | 4.401 | 4.393 | 4.383 | 4.369 |
| 520 | 4.548 | 4.538 | 4.519 | 4.481 | 4.419 | 4.408 | 4.401 | 4.395 |
| 530 | 4.545 | 4.536 | 4.515 | 4.477 | 4.406 | 4.398 | 4.385 | 4.378 |
| 540 | 4.584 | 4.575 | 4.548 | 4.516 | 4.438 | 4.433 | 4.420 | 4.411 |
| 550 | 4.593 | 4.585 | 4.565 | 4.528 | 4.453 | 4.446 | 4.434 | 4.424 |
| 560 | 4.597 | 4.587 | 4.571 | 4.533 | 4.460 | 4.452 | 4.437 | 4.430 |
| 570 | 4.589 | 4.576 | 4.556 | 4.522 | 4.446 | 4.433 | 4.425 | 4.413 |
| 580 | 4.564 | 4.549 | 4.534 | 4.496 | 4.420 | 4.412 | 4.400 | 4.390 |
| 590 | 4.573 | 4.566 | 4.549 | 4.510 | 4.437 | 4.430 | 4.419 | 4.413 |
| 600 | 4.592 | 4.587 | 4.565 | 4.526 | 4.447 | 4.440 | 4.433 | 4.425 |
| 610 | 4.574 | 4.561 | 4.546 | 4.513 | 4.430 | 4.424 | 4.417 | 4.409 |
| 620 | 4.586 | 4.573 | 4.556 | 4.518 | 4.443 | 4.433 | 4.420 | 4.414 |
| 630 | 4.596 | 4.585 | 4.568 | 4.529 | 4.453 | 4.444 | 4.434 | 4.427 |
| 640 | 4.593 | 4.584 | 4.566 | 4.526 | 4.455 | 4.447 | 4.434 | 4.430 |
| 650 | 4.581 | 4.573 | 4.553 | 4.524 | 4.434 | 4.432 | 4.428 | 4.421 |
| 660 | 4.559 | 4.551 | 4.531 | 4.492 | 4.418 | 4.410 | 4.399 | 4.393 |
| 670 | 4.555 | 4.542 | 4.518 | 4.488 | 4.407 | 4.400 | 4.389 | 4.376 |
| 680 | 4.546 | 4.543 | 4.519 | 4.492 | 4.405 | 4.389 | 4.387 | 4.371 |
| 690 | 4.541 | 4.527 | 4.507 | 4.479 | 4.408 | 4.387 | 4.386 | 4.365 |
| 700 | 4.536 | 4.530 | 4.503 | 4.485 | 4.403 | 4.389 | 4.377 | 4.375 |

Table 2 Birefringence $-(n_e - n_0) \times 10^3$ of muscovite mica plate sample 2 of thickness 33.1 μm from 298K to 358K.

| λ (nm) | 298K | 308K | 318K | 328K | 338K | 348K | 358K |
|---|---|---|---|---|---|---|---|
| 300 | 4.346 | 4.338 | 4.308 | 4.286 | 4.281 | 4.274 | 4.274 |
| 310 | 4.431 | 4.419 | 4.389 | 4.374 | 4.368 | 4.360 | 4.359 |
| 320 | 4.472 | 4.457 | 4.430 | 4.411 | 4.405 | 4.397 | 4.397 |
| 330 | 4.495 | 4.479 | 4.450 | 4.434 | 4.429 | 4.423 | 4.421 |
| 340 | 4.510 | 4.491 | 4.463 | 4.450 | 4.442 | 4.436 | 4.433 |
| 350 | 4.521 | 4.503 | 4.476 | 4.461 | 4.454 | 4.449 | 4.446 |
| 360 | 4.529 | 4.512 | 4.485 | 4.470 | 4.463 | 4.458 | 4.456 |
| 370 | 4.541 | 4.520 | 4.495 | 4.482 | 4.473 | 4.467 | 4.465 |
| 380 | 4.553 | 4.529 | 4.507 | 4.490 | 4.486 | 4.481 | 4.474 |
| 390 | 4.558 | 4.542 | 4.518 | 4.502 | 4.495 | 4.491 | 4.485 |
| 400 | 4.567 | 4.550 | 4.528 | 4.514 | 4.507 | 4.502 | 4.496 |
| 410 | 4.576 | 4.558 | 4.533 | 4.522 | 4.514 | 4.507 | 4.501 |
| 420 | 4.585 | 4.563 | 4.541 | 4.529 | 4.518 | 4.515 | 4.511 |
| 430 | 4.595 | 4.568 | 4.548 | 4.535 | 4.524 | 4.521 | 4.514 |
| 440 | 4.599 | 4.575 | 4.554 | 4.542 | 4.534 | 4.530 | 4.522 |
| 450 | 4.605 | 4.577 | 4.555 | 4.541 | 4.534 | 4.530 | 4.522 |
| 460 | 4.611 | 4.587 | 4.564 | 4.552 | 4.544 | 4.540 | 4.536 |
| 470 | 4.612 | 4.586 | 4.569 | 4.553 | 4.543 | 4.540 | 4.533 |
| 480 | 4.611 | 4.591 | 4.577 | 4.559 | 4.558 | 4.549 | 4.544 |
| 490 | 4.607 | 4.603 | 4.581 | 4.574 | 4.565 | 4.558 | 4.551 |
| 500 | 4.616 | 4.611 | 4.590 | 4.580 | 4.568 | 4.567 | 4.560 |
| 510 | 4.647 | 4.593 | 4.581 | 4.560 | 4.557 | 4.552 | 4.548 |
| 520 | 4.610 | 4.602 | 4.586 | 4.574 | 4.569 | 4.566 | 4.557 |
| 530 | 4.664 | 4.597 | 4.577 | 4.568 | 4.560 | 4.557 | 4.547 |
| 540 | 4.664 | 4.632 | 4.612 | 4.595 | 4.590 | 4.584 | 4.572 |
| 550 | 4.659 | 4.640 | 4.626 | 4.613 | 4.605 | 4.600 | 4.586 |
| 560 | 4.654 | 4.647 | 4.630 | 4.615 | 4.609 | 4.605 | 4.593 |
| 570 | 4.677 | 4.630 | 4.616 | 4.599 | 4.593 | 4.591 | 4.578 |
| 580 | 4.679 | 4.612 | 4.593 | 4.578 | 4.572 | 4.567 | 4.554 |
| 590 | 4.647 | 4.624 | 4.614 | 4.601 | 4.596 | 4.592 | 4.584 |
| 600 | 4.670 | 4.642 | 4.621 | 4.610 | 4.608 | 4.599 | 4.587 |
| 610 | 4.659 | 4.616 | 4.606 | 4.593 | 4.592 | 4.586 | 4.574 |
| 620 | 4.675 | 4.631 | 4.614 | 4.602 | 4.600 | 4.589 | 4.580 |
| 630 | 4.673 | 4.645 | 4.626 | 4.612 | 4.611 | 4.601 | 4.593 |
| 640 | 4.660 | 4.641 | 4.629 | 4.618 | 4.609 | 4.607 | 4.593 |
| 650 | 4.653 | 4.635 | 4.618 | 4.606 | 4.601 | 4.601 | 4.590 |
| 660 | 4.675 | 4.609 | 4.593 | 4.585 | 4.579 | 4.576 | 4.558 |
| 670 | 4.680 | 4.603 | 4.585 | 4.570 | 4.569 | 4.565 | 4.555 |
| 680 | 4.699 | 4.597 | 4.578 | 4.576 | 4.568 | 4.551 | 4.548 |
| 690 | 4.666 | 4.601 | 4.570 | 4.566 | 4.561 | 4.558 | 4.546 |
| 700 | 4.635 | 4.596 | 4.592 | 4.568 | 4.566 | 4.558 | 4.552 |

## 4.2 Error analysis

The error of the mica plate's birefringence mainly arises from the error in the determination of the mica plate's thickness and the error of the wavelength and phase retardation. From equation (13), the maximum error of $(n_e - n_0)$ is

$$\delta(n_e - n_0) = \left| \frac{\lambda \Delta}{2\pi d_T^2} \delta d \right| + \left| \frac{\Delta}{2\pi d_T} \delta \lambda \right| + \left| \frac{\lambda}{2\pi d_T} \delta \Delta \right| \tag{14}$$

The error of the mica plate's birefringence as to mica plate sample 1 in the visible spectrum is shown in the following.

Since the uncertainty in mica plate's thickness value is 0.2 μm, the maximum value of $\frac{\lambda \Delta}{2\pi d_T^2} \delta d$ is $2.72 \times 10^{-5}$ in the 400–700 nm spectrum (the maximum value of $\lambda \Delta$ is gained at 560nm from our data).

As to the USPME, the accuracy of the wavelength is 0.1 nm. So the maximum value of $\frac{\Delta}{2\pi d} \delta \lambda$ is $0.11 \times 10^{-5}$ (the maximum value of $\Delta$ is gained at 400nm).

Through repeated measurement, we know that the maximum error of $\Delta$ is $\pm 0.1°$, so that the maximum value of $\frac{\lambda}{2\pi d} \delta \Delta$ is $0.58 \times 10^{-5}$ (when $\lambda$ is 700nm) in the 400–700 nm spectrum.

From the above discussion the maximum error of $(n_e - n_0)$ as to mica plate sample 1 in the visible spectrum is less than $3.4 \times 10^{-5}$. Using the same method the maximum error of $(n_e - n_0)$ is less than $3.1 \times 10^{-5}$ in the ultraviolet spectrum.

As to mica plate sample 2, the maximum error of $(n_e - n_0)$ is less than $3.5 \times 10^{-5}$ in the visible spectrum and $3.2 \times 10^{-5}$ in the ultraviolet spectrum.

## 5 Conclusion

Using the standard transmission ellipsometry, we have demonstrated a method to measure the phase retardation and birefringence of muscovite mica plate. For two mica plate samples, we gained the phase retardation and birefringence dispersion in the temperature range of 223K to 293K and 298K to 358K respectively in the spectrum of 300 to 700 nm. From error analysis the data of mica birefringence have an accuracy of better than $3.5 \times 10^{-5}$ in the 300–700 nm spectrum. The phase retardation and birefringence drop down when the temperature rises at the fixed wavelength from the experimental results. The method described above can also be used to measure the phase retardation and birefringence of the other uniaxial or some biaxial crystals.

## Acknowledgement

This study was supported by the National Youth Natural Science Foundation of China (No. 61107030).

**Appendix A:**

We set $T_{ix}$ and $T_{iy}$ as the transmitted light at the $x$ and $y$ direction separately. As to the $x$ direction, the composed light is

$$\tau_x = T_{1x} + T_{2x} + T_{3x} + \cdots = t_x^2 - t_x^2 r_x^2 \exp(-i2\delta_x) + t_x^2 r_x^4 \exp(-i4\delta_x) + \cdots, \quad (A1)$$

where $r_x = \frac{n - n_x}{n + n_x}$, $t_x = \frac{2n}{n + n_x}$, $\delta_x = \frac{2\pi n_x d}{\lambda}$, with $n$ is the refractive index of the incident medium, $n_x$ is the refractive index at the $x$ direction in the mica wave plate, $\lambda$ is the wavelength of the incident light. So the whole composed light at $x$ direction is

$$\tau_x = \frac{t_x^2}{1 + r_x^2 \exp(-i2\delta_x)}. \quad (A2)$$

Using the same deduction method, the whole composed light at $y$ direction is

$$\tau_y = \frac{t_y^2}{1 + r_y^2 e(-i2\delta_y)}. \quad (A3)$$

Considering the multiple reflections, the Jones matrix should be

$$J_m = \begin{bmatrix} \dfrac{t_x^2}{1+r_x^2\exp(-i2\delta_x)} & 0 \\ 0 & \dfrac{t_y^2}{1+r_y^2\exp(-i2\delta_y)} \end{bmatrix} \begin{bmatrix} \exp(-i\delta_x) & 0 \\ 0 & \exp(-i\delta_y) \end{bmatrix}$$

$$= \begin{bmatrix} \dfrac{t_x^2\exp(-i\delta_x)}{1+r_x^2\exp(-i2\delta_x)} & 0 \\ 0 & \dfrac{t_y^2\exp(-i\delta_y)}{1+r_y^2\exp(-i2\delta_y)} \end{bmatrix}. \qquad (A4)$$

Equation (A4) is equation (5). If we make

$$\psi_x\exp(i\Delta_x) = \dfrac{t_x^2\exp(-i\delta_x)}{1+r_x^2\exp(-i2\delta_x)},\ \psi_y\exp(i\Delta_y) = \dfrac{t_y^2\exp(-i\delta_y)}{1+r_y^2\exp(-i2\delta_y)},\ \psi = \dfrac{\psi_y}{\psi_x},\ \Delta = \Delta_y - \Delta_x,\ \text{then}\ J_m$$

can be set as $J_m = \psi_x\exp(i\Delta_x)\begin{bmatrix} 1 & 0 \\ 0 & \psi\exp(i\Delta) \end{bmatrix}.$ \qquad (A5)

Equation (A5) is equation (6). From calculation we gain

$$\begin{cases} \tan\Delta_x = \dfrac{1-r_{1x}^2}{1+r_{1x}^2}\tan\delta_x \\ \tan\Delta_y = \dfrac{1-r_{1y}^2}{1+r_{1y}^2}\tan\delta_y \end{cases}. \qquad (A6)$$

Then

$$tg\Delta = \dfrac{\tan\Delta_y - \tan\Delta_x}{1+\tan\Delta_y\tan\Delta_x} = \dfrac{(1-r_{1y}^2)(1+r_{1x}^2)\tan\delta_y - (1-r_{1x}^2)(1+r_{1y}^2)\tan\delta_x}{(1+r_{1y}^2)(1+r_{1x}^2)+(1-r_{1y}^2)(1-r_{1x}^2)\tan\delta_x\tan\delta_y}, \qquad (A7)$$

So we can gain equation (7): $\Delta = \arctan\left(\dfrac{(1-r_{1y}^2)(1+r_{1x}^2)\tan\delta_y - (1-r_{1x}^2)(1+r_{1y}^2)\tan\delta_x}{(1+r_{1y}^2)(1+r_{1x}^2)+(1-r_{1y}^2)(1-r_{1x}^2)\tan\delta_x\tan\delta_y}\right).$